\documentclass[10pt]{iopart}

\usepackage{iopams}  
\usepackage{graphicx}
\begin{document}

\title[~]{Recent PQCD calculations of heavy quark production}

\author{Ivan Vitev\dag\ 
\footnote[3]{E-mail address: {\tt ivitev@lanl.gov} }
}

\address{\dag\ Los Alamos National Laboratory, Theoretical 
Division and Physics Division, Mail Stop H846, Los Alamos, 
NM 87545, USA}

\begin{abstract}

We summarize the results of a recent study  of heavy quark 
production and attenuation in cold nuclear matter. 
In p+p collisions, we investigate  the relative contribution 
of partonic sub-processes to $D$ meson production and 
$D$ meson-triggered inclusive di-hadrons to lowest 
order in perturbative QCD. While gluon fusion dominates 
the creation of large angle  $D\bar{D}$ pairs, charm on 
light parton scattering determines 
the yield of single inclusive $D$ mesons. The distinctly 
different non-perturbative fragmentation of $c$ quarks into 
$D$ mesons versus the fragmentation of quarks and gluons into 
light hadrons results in a strong transverse momentum 
dependence of anticharm content of the away-side 
charm-triggered jet. In p+A reactions, we calculate and 
resum the coherent nuclear-enhanced power corrections 
from the final-state partonic scattering in the medium.
We find that single and double inclusive open 
charm production can be suppressed as much as the 
yield of neutral pions from  dynamical high-twist 
shadowing. Effects of energy loss in p+A collisions 
are also investigated in the incoherent Bertsch-Gunion limit  
and may lead to significantly weaker transverse momentum 
dependence of the nuclear attenuation.

\end{abstract}


\submitto{\JPG}

\section{Introduction}

A useful probe of the dense nuclear matter created in collisions 
of heavy nuclei at the relativistic heavy ion collider (RHIC)
is one that is sensitive to dynamical scales and can be both 
cleanly measured experimentally and reliably calculated 
theoretically~\cite{Vitev:2006bi}.
Because of color confinement, only hard probes, i.e. those with 
large momentum transfers, can be reliably calculated in 
the perturbation theory of Quantum Chromodynamics 
(QCD). Experimental measurements of inclusive particle 
suppression are now able to test jet quenching theory out to 
transverse momenta as large as $p_T \sim  20$~GeV~\cite{Vitev:2006uc}.  
On the other hand, typical dynamical scales of the nuclear matter
produced in relativistic heavy ion collisions are on the order of 
hundreds of MeV, which is both much smaller than the scale of a 
hard probe and  non-perturbative. 
Therefore, an ideal probe should be not only ``hard'' but also 
sensitive to this ``soft'' physics.  Open charm production has 
a potential to satisfy these criteria because of the two 
distinctive scales of the open charm meson:
the charm quark mass, $m_c \sim 1.5$~GeV, a relatively hard scale, 
and the binding energy, $\sim M_D-m_c\sim$ hundreds MeV,  
which may be relevant to the fragmentation and dissociation of 
$D$ mesons.

With this in mind, we  present a baseline study of  
heavy quark production and the first results on 
heavy-quark-triggered large-angle correlations in elementary
nucleon-nucleon reactions~\cite{Vitev:2006bi}. Next, we 
investigate the many-body QCD effects in single- and 
double-inclusive open  charm production. 
Because of the quantitative importance of multiple scattering 
for inferring properties of the nuclear matter, the models 
of in-medium interactions should be subjected to experimental 
verification for heavy-flavor production. One may begin to 
gain confidence by applying the theoretical description 
of such effects to heavy ion collisions in simpler situations, 
such as proton-nucleus (p+A) scattering.  
In  this case,  plasma properties are not an issue 
and the main medium effect is, therefore, due to the multiple 
interactions in cold nuclei.
The medium effects that we plan to investigate are also
present in nucleus-nucleus (A+A) reactions during the interaction
time $\tau_{int.} = 2R_A/\gamma \ll \tau_{eq.}  \ll \tau_{QGP}$
and cannot be neglected. 
Here $R_A$ is the nuclear radius, $\gamma$ is the Lorentz gamma
factor of the nucleus, $\tau_{eq.}$ is the equilibration time
and  $\tau_{QGP}$ is the lifetime of the plasma.

We note that full technical details for the results presented 
in these proceedings are given in our complete 
manuscript~\cite{Vitev:2006bi}.

\section{$D$ meson production and correlations in 
lowest order PQCD}

A formal expansion for the differential cross 
section of heavy quark production can be written 
as follows:
\begin{eqnarray}
\hspace*{-1cm}
 \frac{d\sigma}{dy d^2 p_T}
& = & \left( \alpha_s^2 A(m,p_T)  + 
\alpha_s^3 B(m,p_T)  + \cdots  \right) + \left( \alpha_s^2  
\sum_{i=2}^\infty  a_i (\alpha_s \ln \mu/m)^i  \right.  \nonumber  \\
&&  \left. +  \alpha_s^3  
\sum_{i=1}^\infty b_i (\alpha_s \ln \mu/m)^i + \cdots \right) 
G(m, p_T) + \cdots \;, \qquad \qquad
\label{formal}
\end{eqnarray}
where $G(m, p_T) \rightarrow 1$ when  $m/p_T \rightarrow 0$,
and power suppressed terms $\sim (m/p_T)^n$ are not shown.
In Eq.~(\ref{formal}) $A, B, a_i, b_i$ are coefficient functions
and logarithms arise  due to the new mass 
scale, $m$. These are known to next-to-leading logarithm  
(NLL), but numerically,  their contribution to the cross section is small 
up to $p_T \sim 50$~GeV~\cite{Cacciari:2005rk}. 
Existing calculations
of charm and bottom production~\cite{Cacciari:2005rk} treat
quarks as heavy, assuming  $\phi_{c,b/N}(x,\mu_f)\equiv 0$.     
We here evaluate the differential cross sections for 
open charm production and  
open-charm-triggered di-hadron correlations in the Born 
approximation.  We include the charm contribution
from the nucleon wave-function explicitly~\cite{Vitev:2006bi}, 
since this 
approach leads to a faster convergence of the perturbation 
series~\cite{Olness:1997yc} and correspondingly smaller 
next-to-leading (NLO) order K-factors.

\begin{figure}[t!]
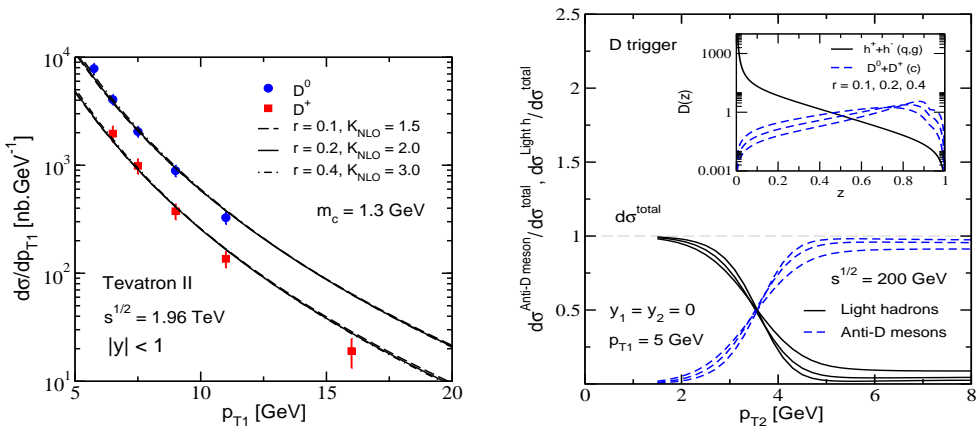

\includegraphics[width=2.4in,height=2.in]{TeVII.eps}
\hspace*{0.5cm}
\includegraphics[width=2.4in,height=2.2in]{D-frag1.eps} 
\caption{ Left panel from~\cite{Vitev:2006bi}: 
cross section for $D^0$ and $D^+$
charm meson production to LO in perturbative QCD at
$\sqrt{s}=1.96$~TeV. Data is from CDF~\cite{Acosta:2003ax}.
Right panel from~\cite{Vitev:2006bi}: 
 contribution of light hadrons and anticharm mesons to the 
{\em fractional production cross section} for 
$p_{T_1} = 5$~GeV $D^0+D^+$ meson-triggered away-side
correlations for  $y_1 = y_2 = 0$. The insert
illustrates the difference in the fragmentation of
light quarks and gluons into hadrons versus that for the 
$c$ quark into charm mesons.  }
\label{ppDtrig}
\end{figure}

There is a direct relation between the absolute normalization
of the cross section and the  hardness of the non-perturbative 
$D$ meson fragmentation. 
We compare in the left panel of Fig.~\ref{ppDtrig} 
the calculated $D^0$ and $D^+$ cross sections to the 
Tevatron Run~II $\sqrt{s}= 1.96$~TeV data~\cite{Acosta:2003ax}.  
We find that three different combinations  
($K_{NLO}  = 1.5$, $r = 0.1$), ($K_{NLO}  = 2$, $r = 0.2$) and
 ($K_{NLO}  = 3$, $r = 0.4$) yield little difference in 
the  single inclusive charm meson spectrum. We note that 
for the same choice of the fragmentation parameter $r=0.1$, 
a LO calculation with standard $\phi_{c/N}(x,\mu_f)\neq 0$  gives
open charm cross sections similar to the ones from a NLO 
calculation that treats flavor as ``heavy''.

Besides the inclusive spectra, we show our first results
on charm meson-triggered away-side di-hadrons.
 The expectation for non-trivial $p_{T_2}$ dependence of  
such large-angle correlations is based on the 
very different  behavior with respect to  
$z = p^{hadron}/p^{parton}$  
of the fragmentation functions of partons into light 
hadrons  compared to those of heavy quarks 
into charm and beauty meson~\cite{Vitev:2006bi}.  
The insert in the right panel of Fig.~\ref{ppDtrig} 
shows that while light hadrons  favor 
soft decays of their parent partons, heavy quark fragmentation is 
very hard. For $z \sim 0.6 - 0.9$ there is more than an 
order of magnitude enhancement in the
decay probabilities $D_{D^0/c}(z)+D_{D^+/c}(z)$ relative to 
$(1/N)\sum_{i=1,N} (D_{\pi^\pm/i} + D_{K^\pm/i} + D_{p(\bar{p})/i})$,
where $i$ runs over the light and charm quarks, antiquarks and 
the gluon.

Triggering on a $D$ meson fixes the momentum of the charm quark 
much more reliably than does triggering on a light 
hadron~\cite{Qiu:2004da}
in studying away-side correlations. The non-perturbative 
fragmentation, therefore, controls the yields of di-hadrons 
versus the associated momentum $p_{T_2}$ and the abundances 
of the different particle species.  The right panel of 
Fig.~\ref{ppDtrig} shows our prediction  for the hadronic 
composition of the ${D}^{0} + D^{+}$-triggered
away-side jet. At transverse momenta significantly smaller 
than the trigger transverse momentum, $p_{T_2} \ll p_{T_1}$, 
the away-side jet is dominated by pions, kaons and protons. 
At transverse momenta  $p_{T_1} \simeq p_{T_2}$, the %
away-side jet is expected to be dominated almost completely 
by $\bar{D}^{0}$ and $D^{-}$ mesons. Since there is little 
sensitivity to the choice of $r$, our prediction is robust 
and can be used as a test of the production mechanism
of heavy quarks.

\section{Resummed QCD power corrections to open
         charm production}

\begin{figure}[t!]
\includegraphics[width=2.6in,height=1.in]{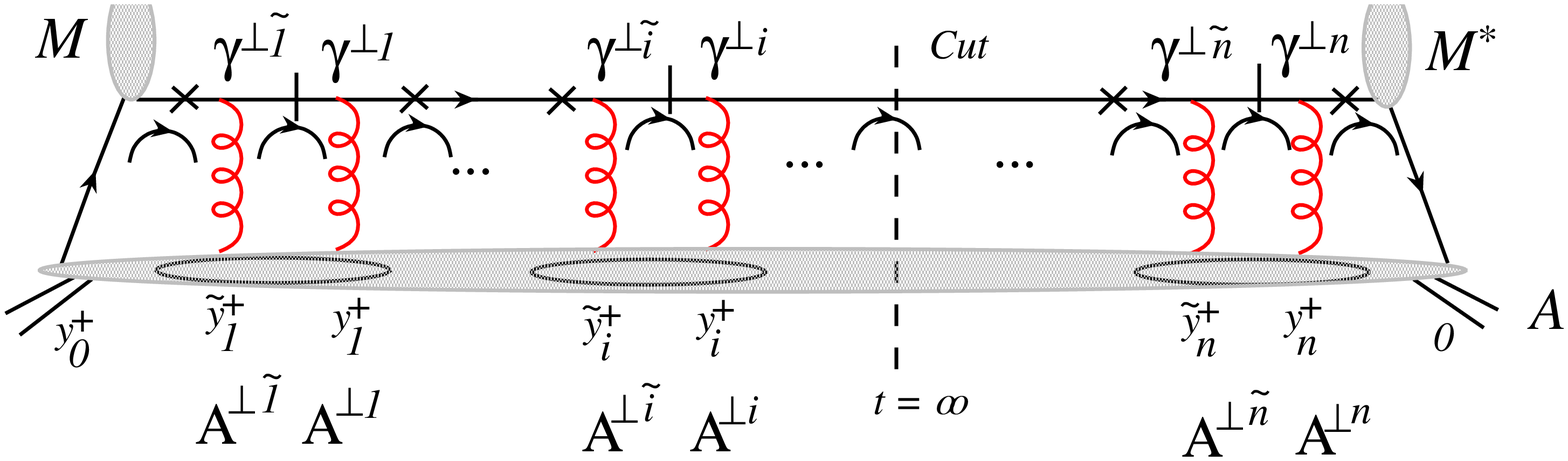}
\hspace*{0.5cm}
\includegraphics[width=2.2in,height=1.8in]{example.eps} 
\caption{ Left panel from~\cite{Vitev:2006bi}: 
multiple coherent scattering of the outgoing 
partons in proton-nucleus reactions in the $t$-channel. We have 
denoted by $ \longrightarrow  
\hspace*{-0.5cm} \times \hspace*{0.25cm}$ and 
$ \longrightarrow  \hspace*{-0.32cm} \mid \hspace*{0.3cm} $ 
the long distance and contact propagators, respectively. 
We have indicated the lightcone positions, e.g. $y_i^+$.
Arcs show the momentum routing.
Right panel from~\cite{Vitev:2006bi}: 
 effects of rescaling of the momentum fraction
$x$ illustrated on the example of the lowest order CTEQ6 parton     
distribution functions for valence and sea u-quarks. Gluons  %
rescatter in the final-state to lowest order only in hadronic 
collisions and are included for completeness.   }
\label{DIS}
\end{figure}

In the case of hadronic collisions there is indeed  similarity 
with the DIS dynamics in the final-state rescattering 
of the struck, small $x_b$, parton from the nucleus, as shown 
in the left panel of Fig.~\ref{DIS}. Such nuclear-enhanced 
power corrections are equivalent to dynamical mass for the 
interacting quark, similar to the generation of dynamical 
electron mass in a strong magnetic field, 
\begin{equation}
m_{dyn}^2 =  \xi^2 A^{1/3}  =  
   A^{1/3} \frac{3 \pi \alpha_s}{8\, r_0^2}  
\lim_{x\rightarrow 0}\frac{1}{2} xG(x) \;.
\label{xi2} 
\label{dyn}
\end{equation}
They lead to suppression of the single- and double-inclusive hadron 
production cross sections in deep inelastic scattering 
(DIS)~\cite{Qiu:2003vd} and forward rapidity p+A 
reactions~\cite{Qiu:2004da},  as long as the 
minimum coherence criterion  $x < 0.1$ is satisfied.

The mechanism behind the reduction of the deep inelastic
inclusive scattering cross section and the differential hadron 
production cross sections is illustrated  schematically 
in the right panel of Fig.~\ref{DIS}. 
PDFs are falling functions of the momentum fraction $x$. 
The evaluation of the parton  distribution function at an 
effectively larger value of $x$  leads to a 
reduction of the cross section. 
From our theory we can easily derive a hierarchy of  dynamical 
high-twist  shadowing, 
\begin{eqnarray}
\hspace*{-1cm}S(x,Q^2) = \left|\; \frac{  - \Delta \sigma (x,Q^2) }
          {  \sigma (x, Q^2) } \; \right| \; , &&  
\quad S_v (x,Q^2) < S_s(x,Q^2) < S_g(x,Q^2)  \;,     
\label{hier}
\end{eqnarray}
for cross sections with multiple scattering dominated by
valence quarks, sea quarks or gluons, respectively.

In~\cite{Vitev:2006bi}, we completed the summation of 
all-twist, nuclear-enhanced power corrections for massive-quark 
final states. In hadronic collisions, we identify 
the function that contains the small
$x_b$ dependence $ F_{ab\rightarrow cd}(x_b) = {\phi_{b/N}(x_b) 
|M_{ab\rightarrow cd}|^2}/{x_b}$.
Implementing  the normalization  to  no nuclear effect 
on a single nucleon as follows, $A^{1/3} \rightarrow A^{1/3} -1 $,  
and having allowed for both massless and massive partons 
(quarks or gluons) we find~\cite{Vitev:2006bi}  
\begin{eqnarray}
 F_{ab\rightarrow cd}(x_b) & \Rightarrow & 
 F_{ab\rightarrow cd}\left( x_b\left[ 1+ C_d \frac{\xi^2 (A^{1/3}-1)}
{-\hat{t} + m_d^2}  \right] \right)\;.
\label{resumt}
\end{eqnarray}
In Eq.~(\ref{resumt}) $C_d = 1\; (9/4)$ for quarks (gluons), 
respectively. When $ m_d \rightarrow 0 $  we recover the result 
of~\cite{Qiu:2004da}. The mass here reduces the effect of power
corrections at $-\hat{t} \simeq  m_d^2$.

While we have focused in detail on the  power corrections 
in the $t$-channel, it should be noted that the coherent initial- 
and final-state interactions of the incoming parton from the 
proton  have also been calculated~\cite{Vitev:2006bi}. 
The results, that we present here, are 
analogous to Eq.~(\ref{resumt}), 
\begin{eqnarray}
\label{resums}
 F_{ab\rightarrow cd}(x_b) & \Rightarrow &  
 F_{ab\rightarrow cd}\left(x_b \left[ 1+ C_a
\frac{\xi^2(A^{1/3}-1)}{-\hat{s}}  \right]\right)  \;, \\
 F_{ab\rightarrow cd}(x_b) & \Rightarrow & 
  F_{ab\rightarrow cd}\left(x_b \left[ 1+ C_c
\frac{\xi^2(A^{1/3}-1) }{-\hat{u}+m_c^2}  \right]\right)  \;. 
\label{resumu}
\end{eqnarray}
By direct comparison  
of the nuclear-$A^{1/3}$-enhanced power corrections in 
Eqs.~(\ref{resumt}), (\ref{resums}) and (\ref{resumu}) 
it is easy to see that in the forward rapidity region  
$|\hat{t}| \ll |\hat{s}|,  |\hat{u}|$ only DIS-like 
dynamical shadowing corrections are important.

In processes with no strong final-state 
interactions, such as the Drell-Yan, coherent multiple
initial-state scattering can give only {\em  enhancement} 
of the cross section, as shown in Eq.~(\ref{resums}), because 
of the rescaling of $x_b$ to smaller versus larger values. 
Thus, initial-state multiple soft 
interactions without energy loss always lead to a Cronin 
effect,  independently of whether 
they are treated as coherent or not. The derivations, 
summarized here, are critical to elucidating the dynamical 
origin of nuclear effects in high energy hadronic reactions.  
We have shown, through explicit calculation, that such
effects are process dependent and may change their 
sign. Therefore, these are {\em not} factorizable as
a part of the parton distribution functions and fragmentation
functions.

\section{Energy loss in  cold nuclear matter}

In the presence of a nucleus, 
final-state rescattering of the struck small-$x$ parton with 
its remnants exhausts the similarity 
between hadronic collisions, such as p+A and A+A,  and
DIS~\cite{Qiu:2004da}. 
In $\ell + A$ (DIS), the multiple interactions of the        
incoming  leptons are suppressed by powers of
$\alpha_{em}/\alpha_s$ relative to the struck parton 
scattering.  In contrast, in p+A and A+A the initial- 
and final-state scattering of the incoming  quarks 
and gluons are equally strong. Nuclear modification, 
in particular jet energy loss associated with the 
suppression of particle production, cannot 
be neglected.

To illustrate the importance of energy loss, we study the 
nuclear modification of hadron production over a large range
of center of mass energies and momentum fractions $x_b$. 
We implement the presently well known incoherent limit of 
energy loss for asymptotic $t= - \infty$ on-shell partons 
first derived in~\cite{Gunion:1981qs}. The double differential 
gluon intensity spectrum per scattering and the fractional 
energy loss $\epsilon$  read:   
\begin{equation}
\frac{\omega dN^{(1)}_g}{d \omega d^2{\bf k}}
\propto  \frac{\alpha_s}{\pi^2}
\frac{{\bf q}^2} {{\bf k}^2({\bf k} - {\bf q})^2} \; ,
\quad \epsilon = \frac{\Delta E}{E} \propto \frac{L}{\lambda} 
= \kappa A^{1/3}  \;,  
\label{BGrad}
\end{equation}
where ${\bf k}$ is the transverse momentum of the radiative gluon, 
${\bf q}$ is the transverse momentum transfer from the medium and
$\omega$ is the gluon energy. In Eq.~(\ref{BGrad})
$L/\lambda$ is the number of interactions and 
 $\kappa = 0.0175$  for minimum bias reactions implies that an 
average parton loses $\sim 10\%$
of its energy in a large nucleus such as Au or Pb.

\begin{figure}[t!]
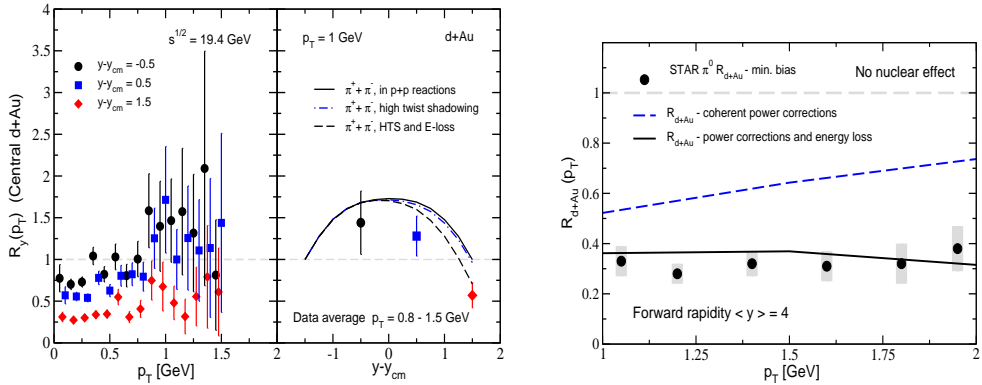

\includegraphics[width=2.6in,height=2.in]{d+Au.eps}
\hspace*{0.5cm}
\includegraphics[width=2.2in,height=1.8in]{STAR-y4.eps} 
\caption{ Left panel from~\cite{Vitev:2006bi}: 
nuclear modification at three different
rapidities $y-y_{cm}$ in $\sqrt{s_{NN}}=19.4$~GeV  d+Au 
collisions. Data is from NA35~\cite{Alber:1997sn}.
Right panel from~\cite{Vitev:2006bi}: calculations of dynamical 
shadowing, with and without cold nuclear matter energy loss, are
compared to the measured~\cite{Adams:2006uz} $\pi^0$ suppression 
at $y=4 $ for d+Au collisions at $\sqrt{s_{NN}}=200$~GeV.
}
\label{eloss}
\end{figure}

We first examine the results obtained by 
the CERN NA35 fixed target experiment with  $y_{cm}=3$.  It 
measured hadron production in d+Au reactions at 
$\sqrt{s_{NN}}=19.4$~GeV~\cite{Alber:1997sn}. 
We take $R_y(p_T)$,  the ratio of hadron spectra 
in different rapidity bins, and $R_{AA}(p_T)$,  
\begin{equation} 
\hspace*{-2.5cm}
R^{AB}_y(p_T) = \frac{ {d \sigma^{h}_{AB}(y,p_T)}/{dy d^2p_T} }   
{ {d \sigma^{h}_{AB}(y_{\rm base},p_T)}/{dy d^2p_T} }   \;,
\;\;   R_{AA}(p_T) = \frac{ {d N^{h}_{AB}(y,p_T)} 
/ {dy d^2p_T} }  {  { T_{AB}(b) d \sigma^{h}_{NN}(y,p_T)} 
/ {dy d^2p_T} }   \;,
\label{Ry}
\end{equation}  
as measures of nuclear matter effects. 
The left panel of Fig.~\ref{eloss} shows this forward 
rapidity suppression of negative hadrons 
relative to the baseline $y-y_{cm} \approx -1.5$ production 
cross section. Dynamical shadowing calculations~\cite{Qiu:2004da} 
give  $ < 5\%$ effect at this energy in this rapidity range.
Conversely, the implementation of energy loss in cold 
nuclear matter leads to much larger suppression 
at forward rapidity and significant improvement 
in the theoretical description of the data in 
Fig.~\ref{eloss}.

In the other extreme,  at $\sqrt{s}=200$~GeV d+Au collisions at 
RHIC, the STAR collaboration observed a factor of 3 suppression 
of $\pi^0$ production at rapidity  $y = 4$~\cite{Adams:2006uz}. 
The right panel of Fig.~\ref{eloss} shows that dynamical high-twist
shadowing, constrained by the DIS data down to values of 
Bjorken $x_B \sim 10^{-4}$, underpredicts the nuclear 
attenuation  by a factor of 2. Incorporating the same energy 
loss that we identified at much smaller center of mass energies, 
we observe very good agreement between theory and the 
data~\cite{Vitev:2006bi}.

\section{Numerical results}

\begin{figure}[t!]
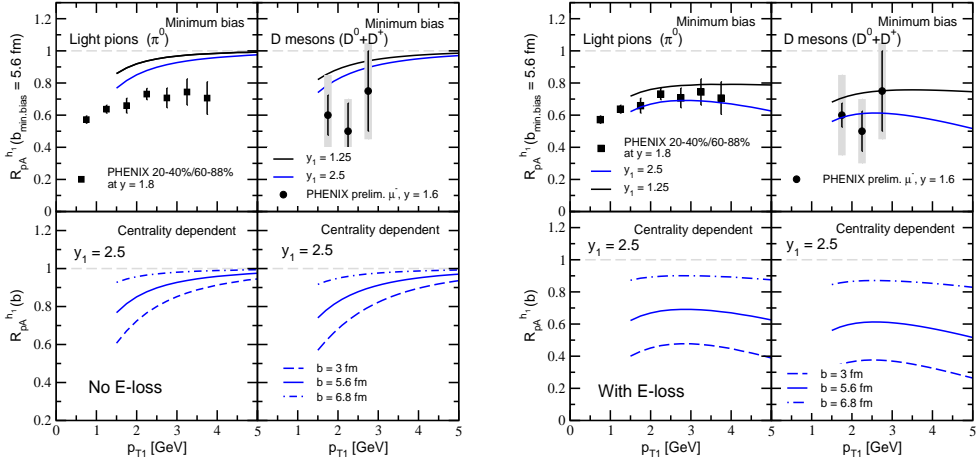

\includegraphics[width=2.4in,height=2.4in]{Power-Charm-NEW.eps}
\hspace*{0.5cm}
\includegraphics[width=2.4in,height=2.4in]{Power-Charm-EL-NEW.eps} 
\caption{ Left panel from~\cite{Vitev:2006bi}: 
suppression of the single inclusive 
hadron production rates in d+Au reactions versus $p_{T_1}$ for 
rapidities $y_1 = 1.25$ (smaller attenuation) and $y_1 = 2.5$ 
(larger attenuation). Theory does not include energy loss.
Right panel from~\cite{Vitev:2006bi}: 
the same calculation with cold nuclear matter energy loss of
the Brtsch-Gunion type~\cite{Gunion:1981qs} included. 
Data is from PHENIX~\cite{Adler:2004eh,Wang:2006de}. Bottom
panels show impact parameter dependence of the calculated  
nuclear modification for central, $b=3$~fm, minimum bias, 
$b_{\rm min.bias}=5.6$~fm and peripheral, $b=6.8$~fm, collisions.
}
\label{EnoE}
\end{figure}

To illustrate the similarities and differences between 
massless and massive final-state partons, we carry out a 
comparative study of the effect of power corrections on single
and double inclusive $\pi^0$ and $D$ meson production. 
The left panel of Fig.~\ref{EnoE} shows the 
suppression of the low and moderate $p_{T_1}$ neutral 
pion cross section  from high-twist shadowing 
relative to the binary scaled p+p 
result. The nuclear modification 
factor is shown for two different forward rapidities, 
$y_1 = 1.25, 2.5$, and three  different centralities, 
$b=3$~fm, $b_{\rm min.bias}=5.6$~fm, $b=6.8$~fm, in  
$\sqrt{s}=200$~GeV d+Au collisions at RHIC. 
At transverse momenta $p_{T_1} = 1.5$~GeV the suppression 
can be substantial, but disappears toward higher $p_{T_1}$ 
due to the power law nature of the effect. 
Coherent power corrections cannot fully  account 
for the nuclear suppression measured by the PHENIX 
experiment~\cite{Adler:2004eh} and the discrepancy 
becomes larger at higher transverse momenta. 
The calculated suppression of $D^0+D^+$ mesons (and 
equivalently, $\bar{D}^0+D^-$ mesons),  in 
deuteron-gold collisions at RHIC is also shown in the 
left panel of Fig.~\ref{EnoE}. It should be noted that
the nuclear modification is very similar to or slightly larger 
than that for $\pi^0$. The reason for this similarity is 
that in both  cases the dominant channel of hadron 
production is via quarks
scattering on gluons, which have the same (large) color singlet 
coupling to the medium, $C_d = 9/4$ in Eq.~(\ref{resumt}). 
In addition, the typical momentum fraction, $z_1$, which enters in 
the determination of the hard scale, $\hat{t} \propto 1 / z_1$, 
is slightly  larger for the $D$ mesons. Preliminary PHENIX data
on the modification of $\mu^-$ coming  from
the decay of heavy flavor at $y = 1.6$ in cold nuclear matter is 
provided as a reference~\cite{Wang:2006de}. Although 
the error bars are large, $D$ mesons seem to be suppressed  as
much as light hadrons.

Having investigated the largest dynamic range
of measurements in proton(deuteron)-nucleus reactions 
accessible to perturbative QCD calculations in the previous
section, we return to 
$D$ meson production and correlations at RHIC. We use
the same fractional energy loss $\epsilon = \Delta E / E$ 
as in Fig.~\ref{eloss}. 
Single inclusive $\pi^0$ and $D^0+D^+$ suppression at rapidities 
$y_1 = 1.25, \, 2.5$ and minimum bias, central and 
peripheral d+Au collisions at RHIC are shown in 
the right panel of Fig.~\ref{EnoE}. In this case, very good agreement
between the QCD theory incorporating cold nuclear matter 
effects and the PHENIX measurement of muons coming from the
decay of light hadrons~\cite{Adler:2004eh} is achieved.    
We find that the magnitude of the $D$ meson suppression 
is similar to that of pions. It is also comparable with
the first forward rapidity results on heavy quark nuclear 
modification in d+Au reactions at RHIC~\cite{Wang:2006de}.  
Nuclear modification of
inclusive two particle production  is seen to follow 
closely that of single inclusive hadrons~\cite{Vitev:2006bi}.

There are similarities  between the calculations that
only include resummed nuclear-enhanced power corrections,  
left panel of Fig.~\ref{EnoE}, and 
the calculations that do not neglect energy loss in cold 
nuclear matter, right panel of Fig.~\ref{EnoE}. 
Both effects are generated  through multiple 
parton scattering in the medium and lead to suppression of 
the rate of hard scattering~\cite{Vitev:2006bi}. In both
 cases single inclusive particle production and large 
angle di-hadron correlations 
are similarly attenuated. Like all nuclear many body effects, 
these increase with the centrality of the collision. 
The difference in the resulting nuclear modification 
is that high-twist shadowing arises from
the coherent final-state scattering of the struck small-$x_b$ 
parton of the nucleus~\cite{Qiu:2003vd} and disappears 
as a function of the transverse momentum~\cite{Qiu:2004da}. 
The energy loss considered here 
arises from the initial-state inelastic scattering of the 
incoming large-$x_a$ parton from the 
proton(deuteron)~\cite{Gunion:1981qs} 
and leads to a suppression which is much more $p_T$ 
independent, similar to final state energy loss
applications~\cite{Vitev:2006uc}.

\section{Conclusions}

In these proceedings, we summarized the results of the first 
perturbative QCD calculation of heavy meson triggered large-angle
di-hadron yields~\cite{Vitev:2006bi} and showed how such 
measurements can provide information on the $D$ meson production 
mechanism and the non-perturbative fragmentation of heavy quarks.
We identified multiple scattering effects that drive the cold 
nuclear matter attenuation of $D$ meson and light hadron 
production at forward rapidities, namely, high twist 
shadowing~\cite{Qiu:2004da,Qiu:2003vd} and parton energy 
loss~\cite{Gunion:1981qs,Gyulassy:2000er}. Our work 
demonstrates~\cite{Vitev:2006bi} how such effects can be 
independently constrained by DIS data (shadowing) and 
low energy p+A measurements~\cite{Alber:1997sn} where 
coherent power corrections are not important (energy loss). 
In the framework of the established QCD collinear factorization, 
our approach of incorporating dynamically generated and  
process-dependent nuclear effects is expected to give a more 
reliable description of particle 
production~\cite{Adams:2006uz,Adler:2004eh,Wang:2006de}  
in p+A collisions. 
This work also provides the baseline for precision QGP tomography 
using heavy quarks as probes of the plasma away from 
midrapidity~\cite{Vitev:2006bi}  where new experimental 
capabilities at RHIC and the LHC are expected to become available.

\vspace*{.3cm}

\noindent {\bf Acknowledgments: } 
I would like to thank my collaborators Terry Goldman, 
Mikkel Johnson and Jianwei Qiu for useful comments. 
This research is supported in part by the US Department 
of Energy under Contract No. W-7405-ENG-3 and
the J. Robert Oppenheimer Fellowship of the 
Los Alamos National Laboratory.

\end{document}